# The Structure of Cold Dark Matter Halos

Julio F. Navarro[1]

*Steward Observatory, The University of Arizona, Tucson, AZ, 85721, USA.*

Carlos S. Frenk

*Physics Department, University of Durham, Durham DH1 3LE, England.*

Simon D.M. White

*Max Planck Institut für Astrophysik, Karl-Schwarzschild Strasse 1, D-85740, Garching, Germany.*

## ABSTRACT

We use N-body simulations to investigate the structure of dark halos in the standard Cold Dark Matter cosmogony. Halos are excised from simulations of cosmologically representative regions and are resimulated individually at high resolution. We study objects with masses ranging from those of dwarf galaxy halos to those of rich galaxy clusters. The spherically averaged density profiles of all our halos can be fit over two decades in radius by scaling a simple "universal" profile. The characteristic overdensity of a halo, or equivalently its concentration, correlates strongly with halo mass in a way which reflects the mass dependence of the epoch of halo formation. Halo profiles are approximately isothermal over a large range in radii, but are significantly shallower than $r^{-2}$ near the center and steeper than $r^{-2}$ near the virial radius. Matching the observed rotation curves of disk galaxies requires disk mass-to-light ratios to increase systematically with luminosity. Further, it suggests that the halos of bright galaxies depend only weakly on galaxy luminosity and have circular velocities significantly lower than the disk rotation speed. This may explain why luminosity and dynamics are uncorrelated in observed samples of binary galaxies and of satellite/spiral systems. For galaxy clusters, our halo models are consistent both with the presence of giant arcs and with the observed structure of the intracluster medium, and they suggest a simple explanation for the disparate estimates of cluster core radii found by previous authors. Our results also highlight two shortcomings of the CDM model. CDM halos are too concentrated to be consistent with the halo parameters inferred for dwarf irregulars, and the predicted abundance of galaxy halos is larger than the observed abundance of galaxies. The first problem may imply that the core structure of dwarf galaxies was altered by the galaxy formation process, the second that galaxies failed to form (or remain undetected) in many dark halos.

---

[1] Bart J. Bok Fellow.

## 1. Introduction

Analytic calculations and numerical simulations both suggest that the density profiles of dark matter halos may contain useful information regarding the cosmological parameters of the Universe and the power spectrum of initial density fluctuations. For example, the secondary infall models of Gunn & Gott (1972) established that gravitational collapse could lead to the formation of virialized systems with almost isothermal density profiles. This result, further refined by Fillmore & Goldreich (1984) and Bertschinger (1985), suggested a cosmological significance for the observation that the rotation curves of galactic disks are flat.

This analytic work was extended by Hoffman & Shaham (1985) and Hoffman (1988), who pointed out that the structure of halos should also depend on the value of the density parameter $\Omega$ and the power spectrum of initial density fluctuations. Studying scale-free spectra, $P(k) \propto k^n$ ($-3 < n < 4$), these authors concluded that halo density profiles should steepen for larger values of $n$ and lower values of $\Omega$. Approximately flat circular velocity curves were predicted for $-2 < n < -1$ and $\Omega \sim 1$. These conclusions were confirmed by N-body experiments (Frenk et al. 1985, 1988, Quinn et al. 1986, Efstathiou et al. 1988a, Zurek et al. 1988) and supported the then emerging Cold Dark Matter scenario, since in this model the effective slope of the power spectrum on galactic mass scales is $n_{eff} \sim -1.5$ (Blumenthal et al. 1984, Davis et al. 1985).

It has become customary to model virialized halos by isothermal spheres characterized by two parameters; a velocity dispersion and a core radius. For galaxies, the halo velocity dispersion (or its circular velocity, defined by $V_c^2 = GM/r$) is often assumed to be directly proportional to the characteristic velocity of the observed galaxy. This assumption allows observations to be compared directly to the results of cosmological N-body simulations or of analytic models for galaxy formation (see e.g. Frenk et al. 1988, White & Frenk 1991, Cole 1991, Lacey et al. 1993, Kauffmann et al. 1993, Cole et al. 1994). However, this simple hypothesis does not seem to be supported by observation. If more massive halos were indeed associated with faster rotating disks and so with brighter galaxies, a correlation would be expected between the luminosity of binary galaxies and the relative velocity of their components. Similarly, there should be a correlation between the velocity of a satellite galaxy relative to its primary and the rotation velocity of the primary's disk. No such correlations are apparent in existing data (see, e.g. White et al. 1983, Zaritsky et al. 1993). A possible explanation may come from the work of Persic & Salucci (1991), who argue that halo circular velocity is only weakly related to disk rotation speed.

Observational estimates of the core radii of dark halos have also led to conflicting results. Large core radii have been advocated in order to accommodate the contribution of the luminous component to the rotation curves of disk galaxies (Blumenthal et al. 1986, Flores et al. 1993); to account for the shape of the rotation curves of dwarf galaxies (Flores & Primack 1994, Moore 1994); and to reconcile X-ray cluster cooling flow models with observations (Fabian et al. 1991). On the other hand, the giant arcs produced by gravitational lensing of background galaxies by galaxy clusters require cluster core radii to be small (see e.g. the review by Soucail & Mellier 1994).

The very existence of a "core", i.e. a central region where the density is approximately constant, has been challenged by recent high resolution numerical work. N-body simulations of the formation of galactic halos and galaxy cluster halos provide no firm evidence for the existence of a core, beyond that imposed by numerical limitations (Dubinski & Carlberg 1991, Warren et al. 1992, Crone, Evrard & Richstone 1994, Carlberg 1994, Navarro, Frenk & White 1995b). These studies indicate that dark matter halos are *not* well approximated by isothermal spheres, but rather have gently changing logarithmic slopes as in the model proposed by Hernquist (1990) for elliptical galaxies,

$$\rho(r) \propto \frac{1}{r(1 + r/r_s)^3}, \qquad (1)$$

or that proposed by Navarro, Frenk & White (1995b) for X-ray cluster halos,

$$\rho(r) \propto \frac{1}{r(1 + r/r_s)^2}. \qquad (2)$$

These profiles are singular (although the potential and mass converge near the center), and possess a well defined scale where the profile changes shape, the scale radius $r_s$. (We will refer to $r_s$ as a "scale" radius and reserve the term "core" to refer to a central region of constant density.) Near the scale radius the profiles are almost isothermal, so these results are quite consistent with the lower resolution simulations mentioned earlier.

The possibility that dark matter density profiles may diverge like $r^{-1}$ near the center has important consequences for the observational issues discussed above. In this paper, we present the results of a large set of high resolution numerical simulations aimed at understanding the density profiles of dark halos in the standard CDM cosmogony. Although this particular cosmogony has fallen into some disrepute since the COBE measurement of temperature fluctuations in the microwave background radiation, it is still the prime example of the general class of hierarchical clustering theories. Many of the conclusions reached from studying this model are directly applicable to other models of structure formation. Indeed, preliminary results of a study of halos formed in scale-free universes corroborate the conclusions presented below (Navarro *et al.* in preparation).

The study of the structure of dark matter halos is particularly well suited to the N-body methods developed over the past decade. Care is needed, however, to model their central regions reliably since these regions are the most vulnerable to systematic errors introduced by finite resolution effects. It is particularly important to ensure that all halos, irrespective of mass, are simulated with comparable numerical resolution. Halos identified in a *single* cosmological simulation are inevitably resolved to varying degrees since massive systems contain more particles than less massive ones. We circumvent this problem by selecting halos with a wide range of masses from cosmological simulations of large regions and then resimulating them individually at higher resolution. We present details of our numerical procedures in §2, and in §3 we test how their limitations affect the structure of our halos. Section 4 presents results for an ensemble of simulations spanning four orders of magnitude in halo mass. Section 5 discusses some implications of our results, and §6 summarizes our conclusions.

## 2. The Numerical Experiments

We simulate the formation of 19 different systems with scales ranging from those of dwarf galaxies to those of rich clusters. The systems were identified in two large cosmological N-body simulations of a standard $\Omega = 1$ CDM universe carried out with a P$^3$M code (Efstathiou *et al* 1985). These simulations followed the evolution of 262,144 particles in periodic boxes of 360 and 30 Mpc on a side, respectively, and were stopped when the *rms* fluctuation in spheres of radius 16 Mpc was $\sigma_8 \equiv 1/b = 0.63$, where $b$ denotes the usual "bias" parameter. (All physical quantities quoted in this paper assume a Hubble constant of 50 km/s/Mpc.) At this epoch, which we identify with the present, we found collapsed systems of varying mass using a friends-of-friends group finding algorithm with the linking parameter set to 10% of the mean interparticle separation. The centers of mass of these clumps were used as halo centers and the radius of the sphere encompassing a mean overdensity of 200 was calculated in each case. Hereafter, we shall refer to this as the "virial" radius of the system and denote it by $r_{200}$. After sorting the clumps by the mass, $M_{200}$, contained within these spheres, we picked 19 of them covering a range of four orders of magnitude in mass ($\sim 3 \times 10^{11} < M_{200}/M_\odot <\sim 3 \times 10^{15}$). We did not select randomly from this mass range, but chose instead halos grouped roughly into three bins of circular velocity, $\sim 100$ km s$^{-1}$, 250 km s$^{-1}$ and $> 450$ km s$^{-1}$. This gives some indication of the "cosmic scatter" in the properties of halos of a given mass.

The particles in each of the 19 selected systems were then traced back to the initial time, where a box containing all of them was drawn. This box was loaded with $32^3$ particles on a cubic grid which was perturbed using the *same* waves as in the original simulation, together with additional waves to fill out the power spectrum to the Nyquist frequency of the new particle grid. The size, $L_{box}$, of this "high-resolution" box was chosen so that all systems had, at $z=0$, approximately the same number of particles within the virial radius, typically $5,000-10,000$. The gravitational softening was chosen to be one percent of the virial radius, and was kept fixed in physical coordinates. (Note that some of the tests described in §3 have fewer particles; for these runs the gravitational softening was slightly larger.) The initial redshift of each run was chosen so that the median displacement of particles within the "high-resolution" box was smaller than the initial particle gridsize. These choices ensure that all halos were resolved to a similar degree at $z=0$. A summary of the numerical parameters of the simulations is given in Table 1. The parameters listed in this Table are defined throughout the text.

Tidal effects due to distant material are represented by using several thousand massive particles to coarse-sample the region surrounding the "high-resolution" box. This procedure for setting up initial conditions is identical to that described in Navarro, Frenk & White (1995a,b), where further details may be found. All simulations were run with the N-body code described by Navarro & White (1993); a second-order accurate, nearest-neighbor binary-tree code, with individual particle timesteps. Typically, the minimum timestep is between $10^{-4}$ and $10^{-5}$ of a Hubble time, low mass systems requiring more timesteps than more massive ones due to their higher central densities.

## 3. Effects of numerical limitations

As mentioned in §1, numerical limitations can influence the structure of model halos. The number of particles, the gravitational softening, the initial redshift, and the timestep can all, in principle, have a significant effect. It is important to check explicitly that our results are insensitive to the particular choice of numerical parameters we have made. For each of the tests described below we chose one of the least massive and one of the most massive halos in our series, and we resimulated them varying the numerical parameters in a systematic fashion. Our conclusion is that, when chosen carefully, numerical parameters do not affect the structure of halos in well resolved regions, *i.e.* at radii which are larger than the gravitational softening and which enclose a sufficiently large number of particles ($\gtrsim 50$).

### 3.1. The initial redshift

Figures 1a and 1b show the effect of varying the initial redshift, $z_i$, of the simulation for two of our halos, a large halo of mass $M_{200} \sim 10^{15} M_\odot$, corresponding to a rich galaxy cluster, and a small halo of mass $M_{200} \sim 10^{11} M_\odot$, corresponding to a dwarf galaxy. These models had only $22^3$ particles within the "high-resolution" box for CPU economy reasons. The panels show the density profile and the circular velocity as a function of radius. The profiles are spherically averaged over bins containing about 30 particles each.

Three different choices of $z_i$ were tried for the small halo, $1 + z_i = 5.5$, 11.0, and 22.0, and two for the large system, $1 + z_i = 5.5$ and 11.0. The density profiles show little change, although for $1 + z_i = 5.5$ the central density of the small halo appears to be substantially underestimated. Since it is a cumulative quantity, the circular velocity profile is more sensitive to variations near the center. Figure 1b shows that starting at $1 + z_i = 5.5$ results in a reduced circular velocity near the center but, as the starting redshift is increased, the profiles converge to a unique shape so that for $1 + z_i = 11.0$ and 22.0 the two curves are indistinguishable. In the case of the massive system, starting at $1 + z_i = 5.5$ or at 11.0 had no significant effect.

The importance of the initial redshift may be understood by considering the initial displacement field of the particles given by the Zel'dovich approximation. At $1 + z_i = 5.5$, the median particle displacement in the small halo simulation was about 4 times the mean interparticle separation. Thus, the fluctuations are not linear on small scales in the initial conditions and the Zel'dovich approximation breaks down. The excess kinetic energy imparted to the particles prevents the formation of dense clumps in the early stages of the simulation, reducing the maximum density of the final system. The median displacement is less than the mean interparticle separation at $1 + z_i = 22.0$, and less than twice the mean interparticle separation at $1 + z_i = 11.0$. In these cases, use of the Ze'ldovich approximation is justified, and the results converge to a unique solution.

In the case of the massive halo, even at $1 + z_i = 5.5$ the median displacement was only about twice the mean interparticle separation. As a result starting this "late" had a negligible effect on the final profile. We conclude that as long as the initial ratio between the median particle displacement and the mean interparticle separation is of order unity (or less) the results are insensitive to the particular choice of $z_i$. The initial redshifts of all runs (quoted in Table 1) satisfy this condition.

### 3.2. The gravitational softening

The gravitational softening, $h_g$, is included in the N-body equations of motion in order to suppress relaxation effects due to two-body encounters. This is accomplished if $h_g$ is set significantly larger than the impact parameter for a large angle deflection in a typical 2-body encounter, $\sim Gm/\sigma^2$, where $m$ is the mass of a particle and $\sigma$ the velocity dispersion of the system (White 1979). Since $\sigma^2 \approx GM_{200}/2\,r_{200} = GmN_{200}/2\,r_{200}$, then $h_g$ should be larger than $2\,r_{200}/N_{200}$. (Here $N_{200}$ is the total number of particles within the virial radius $r_{200}$ and we have assumed that all particles have the same mass.) In our simulations $N_{200}$ is typically of order a few thousand, so choosing $h_g$ of the order $10^{-2} \times r_{200}$ should be safe.

Figures 2a and 2b show that varying the softening length by up to a factor of 5 (within the constraint mentioned above) has little impact on the final structure of the halos outside about one softening length. Three different values of $h_g$ were tried for the large halo and two for the small system. The figures display the density and circular velocity profiles of the same two halos used in the previous subsection. The number of particles within the "high-resolution" box, however, was increased to $32^3$, and the initial redshift was also fixed at $1 + z_i = 22.0$ in both cases.

Note that although $h_g < 10^{-2} \times r_{200}$ could be used, too small a value of $h_g$ would be counterproductive. Smaller softenings require smaller timesteps for the same accuracy, with a consequent increase in CPU time consumption. For the test runs in Figure 2, the number of timesteps increased roughly as $h_g^{1/2}$. Thus, the experiments illustrated in this figure also show that the final halo structures are independent of the number of timesteps in the simulation, typically over $10,000$ timesteps per run.

Finally, comparison of Figures 1 and 2 shows that increasing the number of particles by a factor of three has no noticeable effect on the structure of the halos, except that larger values of $N$ allow us to determine the structure closer to the halo's center. At radii containing more than $\sim 50$ particles, our results are insensitive to the number of particles. For the particle numbers in these runs, we can reliably probe the structure of halos over approximately two decades in radius, between $r_{200}$ and $10^{-2} \times r_{200}$.

## 4. Results

### 4.1. Density Profiles

Figure 3 shows the density profiles of four dark halos of different mass. Halo mass increases from left to right, from $\sim 3 \times 10^{11} M_\odot$ for the smallest system, to $\sim 3 \times 10^{15} M_\odot$ for the largest. The arrows indicate the gravitational softening in each simulation. In all cases, the virial radius is about two orders of magnitude larger than the softening length. The smooth curves represent fits to the simulation data using a model of the form proposed by Navarro, Frenk & White (1995b),

$$\frac{\rho(r)}{\rho_{crit}} = \frac{\delta_c}{(r/r_s)(1 + r/r_s)^2}, \qquad (3)$$

where $r_s = r_{200}/c$ is a characteristic radius an $\rho_{crit} = 3H^2/8\pi G$ is the critical density ($H$ is the current value of Hubble's constant); $\delta_c$ and $c$ are two dimensionless parameters. Note that $r_{200}$ determines the mass of the halo, $M_{200} = 200\rho_{crit}(4\pi/3)r_{200}^3$, and that $\delta_c$ and $c$ are linked by the requirement that the mean density within $r_{200}$ should be $200 \times \rho_{crit}$. That is,

$$\delta_c = \frac{200}{3} \frac{c^3}{(\ln(1+c) - c/(1+c))}. \qquad (4)$$

We will refer to $\delta_c$ as the characteristic overdensity of the halo, to $r_s$ as its scale radius, and to $c$ as its concentration.

A striking feature of Figure 3 is that the same profile shape provides a very good fit to all the halos even though they span nearly four orders of magnitude in mass. The agreement holds for radii ranging from the gravitational softening length to the virial radius. Halo concentration decreases systematically with increasing mass. This may be clearly seen in Figure 4, where we show two of the profiles of Figure 3 (corresponding to the smallest and largest halos), scaled so that the density is given in units of the critical density and the radius in units of the virial radius. Although we show profiles only for a few halos in Figures 3 and 4, these are by no means special. Eq. (3) fits all our halos almost equally well, irrespective of mass.

### 4.2. Circular Velocity Profiles

The circular velocity profile, $V_c(r) = (GM(r)/r)^{1/2}$, contains the same information as the density profile but is less noisy because it is a cumulative measure of the radial structure of a halo. Circular velocity curves for all 19 systems are shown in Figure 5. They all look rather similar, as expected given the similarity in the shapes of the density profiles. The circular velocity rises near the center, then remains almost constant over an extended region, and finally declines near the virial radius. There is a noticeable trend with mass. Larger systems have circular velocities that continue to rise to a larger fraction of the virial radius than do smaller halos. This is more easily seen in Figure 6, where for two halos we scale circular velocity by the value at $r_{200}$, $V_{200} = (GM_{200}/r_{200})^{1/2} \approx (1/2)(r_{200}/\text{kpc})$ km/s, and radii by $r_{200}$ itself. Again, we choose the least and most massive systems to illustrate the trend. Note (i) that the maximum circular velocity of each halo, $V_{max}$, is larger than $V_{200}$ (by up to 40 per cent); (ii) that the ratio $V_{max}/V_{200}$ is larger for low mass halos; and (iii) that the radius $r_{max}$ at which the peak occurs is a larger fraction of the virial radius for more massive systems. This is an alternative way of expressing the fact that low mass systems are significantly more concentrated than high mass ones (see Figure 4).

The dashed lines in Figure 6 are fits to the circular velocity curve predicted from eq.(3);

$$\left(\frac{V_c(r)}{V_{200}}\right)^2 = \frac{1}{x}\frac{\ln(1+cx)-(cx)/(1+cx)}{\ln(1+c)-c/(1+c)}, \quad (5)$$

where $x = r/r_{200}$ is the radius in units of the virial radius. Note that for this model the circular velocity peaks at $r_{max} \approx 2r_s = 2r_{200}/c$.

The dotted line in Figure 6 is a fit to the circular velocity curve of the small halo using the Hernquist (1990) model. The circular velocity in this model is given by

$$\left(\frac{V_H(r)}{V_{max}}\right)^2 = \frac{4(r/r_{max})}{(1+r/r_{max})^2}. \quad (6)$$

Note that although in the inner regions (for overdensities larger than $\sim 1000$) this model provides as good a fit to the data as our adopted profile (eq. 3), it predicts circular velocities that are too low near the virial radius.

### 4.3. Mass dependence of halo properties

As shown in Figures 4 and 6, low mass CDM halos are more centrally concentrated than high mass ones. This is not surprising because lower mass systems generally collapse at higher redshift, when the mean density of the universe was higher. In the spherical top-hat model, the postcollapse density is a constant multiple of the mean cosmic density at the time of collapse. To understand the relation between halo mass and characteristic density requires a formal definition of the time of formation. This is not straightforward because systems are continually evolving, accreting mass through mergers with satellites and neighboring clumps. One possibility is to define the formation time of a halo as the first time when half of its final mass $M$ was in progenitors with individual masses exceeding some fraction $f$ of $M$. With this definition, the typical formation redshifts of halos of differing mass can be predicted analytically.

Lacey & Cole (1993) show that a randomly chosen mass element from a halo of mass $M$ identified at redshift $z_0$ was part of a progenitor with mass exceeding $fM$ at the earlier redshift $z$ with probability

$$P(>fM, z|M, z_0) = erfc\left(\frac{\delta_0(z-z_0)(1+z_0)}{\sqrt{2(\Delta_0^2(fM)-\Delta_0^2(M))}}\right), \quad (7)$$

where $\Delta_0^2(M)$ is the variance of the linear power spectrum at $z = 0$ when smoothed with a top hat filter enclosing mass $M$ (see, for example, eqs. (6) and (7) of White & Frenk (1991) for a definition), and $\delta_0 = 1.69$ is the usual critical linear overdensity for top-hat collapse. This equation is valid for $z > z_0$ and $f < 1$. The formation redshift $z_{form}(M, f, z_0)$ is then defined by setting $P = 1/2$ in eq. (7). Lacey & Cole (1994) tested this formula against N-body simulations for the case $f = 0.5$ and found excellent agreement.

Having assigned a typical formation redshift to halos of a given mass, the mean density of the universe at that redshift provides a natural scaling parameter. Let us assume that the characteristic overdensity of a halo scales as

$$\delta_c(M, f, z_0) = C(f)(1 + z_{form}(M, f, z_0))^3. \quad (8)$$

Figure 7 shows the characteristic density, $\delta_c$ (obtained by fitting eq. 3), as a function of halo mass expressed in units of the non-linear mass, $M_\star(z_0)$. This characteristic mass is defined by the condition $\Delta(M_\star(z)) = 1.69/(1+z)$; for our adopted normalization of the CDM power spectrum, $M_\star \simeq 3.3 \times 10^{13} M_\odot$ at $z = 0$. The curves in Figure 7 show the predictions of eq. (8) for various choices of $f$. The normalizations have been chosen so that the curves all cross at $M_{200} = M_\star$; this implies $C(f) = 1.5 \times 10^4, 5.4 \times 10^3$, and $2.0 \times 10^3$ for $f = 0.5, 0.1$, and $0.01$, respectively. Note that the shapes of these curves are almost independent of $f$ for $f \ll 1$. The agreement between the mass-density relation predicted by eqn. (8) and our N-body results is fair for $f = 0.5$, and improves for smaller values of $f$. Thus the characteristic density of a halo does indeed seem to reflect its formation time.

Figure 8 shows the correlation between concentration, $c$, and the mass of the halo. Although this figure contains the same information as Figure 7 (since $\delta_c$ and $c$ are related by eqn. 4), it shows explicitly that large halos are significantly less concentrated than small ones. Note that the dependence of concentration on mass is quite weak; $c$ changes by less than a factor of four while $M$ varies by four orders of magnitude. Thus, halos differing in mass by less than, say, one order of magnitude, have density profiles that are almost indistinguishable in the scaled variables used in Figure 4. This confirms the results we presented in Navarro, Frenk & White (1995b), where we found that *galaxy cluster* halos in the CDM cosmogony are well approximated by eq. (3) with $c \approx 5$. The present higher resolution simulations show that such a weak concentration is appropriate only for the largest halos, $M_{200}/M_\star \sim 100$. With hindsight, the weak dependence of $c$ on mass conspired with our use in Navarro, Frenk & White (1995b) of a fixed softening length of about 100 kpc for all systems and of a rather "late" starting redshift, $z_i = 3.74$, to give concentration estimates which are slightly lower than those shown in Figure 8.

Figure 9 shows that the maximum circular velocity, $V_{max}$, can differ by up to 40 percent from the circular velocity at the virial radius, $V_{200}$. The ratio $V_{max}/V_{200}$ is an interesting quantity to consider because many analytic and numerical studies use $V_{200}$ to characterize the statistical properties of the halo population and to compare them with the observed properties of the galaxy population. For example, it is often assumed that the rotation velocity of a galactic disk is identical to $V_{200}$, an assumption that clearly breaks down if the circular velocity of a halo varies with radius as shown in Figures 5 and 6.

Finally, Figure 10 shows the correlation between the maximum circular velocity, $V_{max}$, and the radius, $r_{max}$, at which the circular velocity attains that maximum. Note that for $V_{max} = 220$ km/s, the rotation velocity of the Milky Way, $r_{max}$ is about 50 kpc, larger than the typical optical size of galaxies like the Milky Way. Thus, if the dark halo of the Milky Way resembles our model halos, the observed flat rotation curve must be the result of the dissipational collapse of the luminous component rather than a direct reflection of the structure of the dark halo. We discuss further the implications of these results in the following section.

## 5. Discussion

### 5.1. The shapes of the rotation curves of disk galaxies.

The shapes of the rotation curves of disk galaxies have been the subject of numerous studies (Burstein & Rubin 1985, Whitmore *et al.* 1988, Frenk & Salucci 1989, Persic & Salucci 1991, Casertano & van Gorkom 1991). To first approximation, the rotation velocity is observed to be constant within the luminous radius, but there is good evidence now that the rotation curves of less luminous galaxies tend to be rising whilst those of their brighter counterparts tend to be gently declining. These systematic trends have been confirmed in a recent analysis of the large dataset of Matthewson *et al.* (1992) by Persic & Salucci (1995). They find that the optical radius of a galaxy, $r_{opt}$, and the rotation velocity at $r_{opt}$, $V_{opt}$, are strongly correlated, $r_{opt} \approx 20(V_{opt}/200$ km s$^{-1})^{1.16}$ kpc. (The optical radius is defined as the radius that encloses 83% of the total $B$-band luminosity of the galaxy and corresponds to 3.2 exponential disk scalelengths. For a Freeman disk, $r_{opt}$ corresponds to the 25 mag$_B$/arcsec$^2$ isophotal radius.) The total $B$-band luminosity of a galaxy is also strongly correlated with $V_{opt}$ via the Tully-Fisher relation, $L_B \approx 3.3 \times 10^9 (V_{opt}/100$ km s$^{-1})^{2.7} L_\odot$ (Pierce & Tully 1988). All these correlations exhibit rather small scatter.

We now consider the implications of our results on the structure of dark matter halos for these observations. To compute the rotation curve of a model "disk galaxy" forming within a CDM halo, we need to specify three parameters: the mass of the disk, $M_{disk}$, its exponential scalelength, $r_{disk}$, and the mass of the halo, $M_{200}$. (We ignore the contribution of the galaxy's spheroid in this simplified model.) According to the observed correlations mentioned above, $V_{opt}$ uniquely determines the luminosity and scalelength of the disk. Requiring that our model should also exhibit these correlations then reduces the number of parameters to two and these may be taken to be the disk mass-to-light ratio, $(M/L)_{disk}$, and the halo mass. These two parameters may be fixed by matching the amplitude of the observed rotation curve, $V_{opt}$, and its slope at the optical radius.

To carry this program through, we need to allow for the fact that the halo will respond to disk formation by adjusting slightly according to the mass and radius of the disk. We assume that the disk is assembled slowly and that the halo is adiabatically compressed during this process (Blumenthal *et al.* 1986, Flores *et al.* 1993). In this approximation, the radius, $r$, of each halo mass shell after the assembly of the disk is related to its initial radius, $r_i$, by

$$r \left[ M_{disk}(r) + M_{halo}(r) \right] = r_i M_i(r_i). \qquad (9)$$

Here $M_i(r_i)$ is the mass within radius $r_i$ before disk formation (found by integrating eqn. 3), $M_{disk}(r)$ is the final disk mass within $r$ (assumed to be exponential) and $M_{halo}(r)$ is the final dark matter distribution we wish to calculate. We assume that halo mass shells do not cross during compression, so that $M_{halo}(r) = M_{halo}(r_i) = (1 - \Omega_b)M_i(r_i)$, where $\Omega_b = 0.06$ is the initial baryon fraction and is chosen to agree with primordial nucleosynthesis calculations (Walker et al. 1991).

The results of this procedure are illustrated in Figure 11, where we plot the circular velocity of the adiabatically compressed halo (dashed lines) and the disk+halo or "galaxy" rotation curve (solid lines) for galaxies with $V_{opt}$ equal to 100, 200, and 300 km/s. Each curve is labeled by the $B$-band disk mass-to-light ratio required by our model. The "typical" slopes of observed rotation curves near $r_{opt}$, as given by Persic, Salucci & Stel (1995), are shown as dotted lines. These and the model fits may be seen to be declining for rapidly rotating disks and gently rising for slowly rotating disks.

Note that in order to match the observed rotation curves, our models require that disk mass-to-light ratios increase with luminosity, approximately as $(M/L)_{disk} \propto L^{1/2}$. This is because even after adiabatic compression, the circular velocity of the halo is either rising or flat near the optical radius. Thus, to produce a declining rotation curve, the disk (whose own rotation curve is declining in this region) must be relatively more important in brighter galaxies. Assuming a constant $(M/L)_{disk} \approx 1.2$ for all galaxies would result in nearly flat rotation curves at $r_{opt}$ for all models irrespective of $V_{opt}$. The required increase in $(M/L)_{disk}$ with rotation velocity is consistent with the results of Broeils (1992a) and Salucci, Ashman & Persic (1991), who also required a systematically varying $(M/L)_{disk}$ to fit the rotation curves of spiral galaxies.

An important implication of our models is that as $V_{opt}$ for a galaxy increases from 200 to 300 km/s, $V_{200}$ for the surrounding halo increases only from 150 to 170 km/s. This kind of behaviour was previously noted by Persic and Salucci (1991) and has a number of consequences. It affects the interpretation of dynamical data on binary and satellite galaxies, an issue which we discuss in more detail in §5.2. It may also help with an apparent difficulty in theories where galaxies form by the cooling of baryons within dark matter halos (White & Rees 1978). Assuming galaxy/halo systems to have singular isothermal potentials with circular velocity $V_c$ equal to the rotation velocity of the central disk, White & Frenk (1991) estimated X-ray luminosities of order $10^{42}$ erg/s for bright spiral galaxies, well in excess of the observed upper limits on diffuse emission from bright spiral halos. They noted that their model predicted a typical emission temperature of $T_X = 0.19(V_c/250 \text{ km/s})^2$ keV, suggesting that the emission might be soft enough to have been missed. The halo structure implied by our current models has $V_c$ significantly smaller than $V_{opt}$ over the emitting regions and thus implies even softer emission from the cooling gas.

In our models, the dark matter fraction within $r_{opt}$ increases sharply with decreasing galaxy luminosity, from less than $\sim 70\%$ for galaxies with $V_{opt} = 300$ km/s, to more than $\sim 90\%$ for galaxies with $V_{opt} = 100$ km/s. The fraction of the total mass within the virial radius of the halo represented by the disk also depends strongly on $V_{opt}$ and varies from $\sim 0.07$ for $V_{opt} = 300$ km/s, to less than $\sim 0.01$ for $V_{opt} = 100$ km/s. The value for large galaxies is close to our assumed value for the universal baryon fraction. Thus, if pregalactic material had a uniform baryonic mass fraction to begin with, our models require that the transformation of baryons into stars should have been extremely inefficient in halos with circular velocity below about 100 km/s.

Such a trend of increasing global mass-to-light ratio ($M_{200}/L_B$) with decreasing halo mass is also required in hierarchical clustering models of galaxy formation in order to reconcile the steep mass function of dark halos predicted in these models with the shallow faint-end slope of the observed galaxy luminosity function (see, e.g. White & Rees 1978, Frenk et al. 1988, White & Frenk 1991, Lacey et al. 1993, Kauffmann et al. 1993, Ashman, Salucci and Persic 1993, Cole et al. 1994). Thus a detailed theory for galaxy formation within halos of this type is likely to be more successful than previous semi-analytic models in producing realistic galaxy luminosity functions.

In summary, the observed rotation curves of spirals are consistent with the structure of CDM halos provided that the assembly of the luminous component of galaxies was inefficient in low mass halos. The rotation velocity of bright galaxies is largely determined by the contribution of the disk, and is therefore not a good indicator of the circular velocity of the surrounding halo.

### 5.2. The dynamics of binary galaxies and satellite companions.

A major unresolved issue in studies of the dynamics of binary galaxies and of satellites orbiting around bright spirals is the puzzling lack of correlation between the luminosity of the system and the observed relative velocity difference (White et al. 1983, Zaritsky et al. 1993, Zaritsky & White 1994). The puzzle arises because it is commonly assumed that the rotation velocity of galactic disks is a good indicator of the mean circular velocity of their surrounding halos. Since brighter galaxies have more rapidly rotating disks, a strong correlation is expected between the luminosity of a pair and their velocity difference, as well as between the rotation velocity of a primary galaxy and the orbital velocity of its satellites. Observational datasets rule out such correlations with high significance. Indeed, the data seem to favor models where galaxies have extended halos with a mass which depends only very weakly on their luminosity.

The discussion in §5.1 provides an interesting clue to this puzzle. As can be seen in Figure 11, the mean circular velocity of halos surrounding galaxies with rotation velocities larger than about 200 km/s is almost uncorrelated with $V_{opt}$ and, consequently, with the luminosity of the galaxy. (Bright galaxies with $V_{opt}$ of the order or larger than 200 km/s typically constitute the bulk of the systems considered in observational surveys of binaries and satellite/primary systems.) According to the analysis presented above, bright galaxies are typically surrounded by a halo with mean circular velocity $V_{200} \sim 160$ km/s. The mass of such halo is $M_{200} \sim 1.9 \times 10^{12} M_\odot$ within the virial radius $r_{200} \sim 320$ kpc. This agrees with the estimates of Zaritsky & White (1994), who found from their study of the dynamics of satellites surrounding bright spirals that the average circular velocity is between 180 and 200 km/s at 300 kpc. This agreement is encouraging, since the halo properties we infer here are chosen to match the shape of the disk rotation curves, and do not use any information about dynamics at larger radii. The structure of the dark halos which surround bright spiral galaxies seems to be similar to that implied by our models.

### 5.3. The cores of dwarf galaxies.

The luminous component typically constitutes only a small fraction of the total mass within the luminous radius of a dwarf galaxy. Therefore, measurements of the internal dynamics of these dark matter dominated systems are a direct probe of the inner regions of dark matter halos. High-quality rotation curves of several dwarf galaxies indicate that the dark halo circular velocity rises almost linearly with radius over the luminous regions of these galaxies (Carignan & Freeman 1988, Carignan & Beaulieu 1989, Broeils 1992a,b, Jobin & Carignan 1990, Lake et al. 1990). For spherical symmetry, this implies that the halo has a well defined core within which the dark matter density is approximately constant. As pointed out by Moore (1994) and Flores & Primack (1994), this result is inconsistent with the singular halo models favoured by N-body simulations such as those presented in this paper.

Figure 12 shows the contribution of the dark halo to the observed rotation curve of four dwarf galaxies (solid lines). The two curves are meant to encompass the halo contributions allowed by the observations, and correspond to the results of assuming a "maximal disk" or a "maximal halo". These fits *assume* that the halo structure is of the form $\rho(r) = \rho_0/(1 + (r/r_{core})^2)$, and are plotted only in the radial range where the rotation curve is measured. The parameters for each galaxy have been taken from Carignan & Freeman (1988, DDO154), Broeils (1992a, DDO168 and NGC3109), and Lake et al. (1990, DDO170). The dotted lines in these plots are the expected contribution of a CDM dark halo, constrained to match the circular velocity at the outermost radius for which the rotation velocity has been measured. The two dotted lines are meant to represent the most and least concentrated halo compatible with this constraint and with the scatter in the correlations shown in Figure 10 (a factor of $\sim 2$ in $r_{max}$ for a given choice of $V_{max}$).

Figure 12 shows that, except perhaps for DDO170, where observations constrain the halo parameters very poorly, the CDM halos appear too concentrated to fit the observations. We thus agree with the conclusions of Moore (1994) and Flores & Primack (1994). The discrepancy, however, is less dramatic than found by these authors (*e.g.* compare Figure 12 with Figure 1 of Moore 1994). The reason for this difference is that CDM halos are actually less concentrated than assumed by Moore (1994). Moore's figure is based on the simulation of a single CDM halo by Dubinski & Carlberg (1991), who found $r_{max} = 26$ kpc for an object with $V_{max} = 280$ km/s. This contrasts with the mean value of $\sim 60$ kpc found in our simulations for the same circular velocity (see Figure 10). We attribute this difference to the fact that Dubinski & Carlberg stopped their simulations at $z = 1$; their run cannot therefore be considered representative of dark halos identified at $z = 0$.

We conclude that, although the cores of dwarf galaxies pose a significant problem for CDM, the problem is not as bad as previously thought. Perturbations to the central regions of dwarf galaxy halos, resulting perhaps from the sudden loss of a large fraction of the baryonic material after a vigorous bout of star formation (Dekel and Silk 1986), can in principle reconcile the observations of dwarfs with the structure of CDM halos (Navarro, Eke & Frenk 1995).

### 5.4. The abundance of galactic dark halos.

A problem which afflicts all current models of galaxy formation based on the standard CDM cosmogony is related to the high abundance of galactic dark halos (White & Frenk 1991, Kauffmann et al. 1993, Cole et al. 1994). Indeed, dark halos of galactic size are so numerous in this scenario that it is impossible to fit *simultaneously* the observed Tully-Fisher relation and the galaxy luminosity function. This can be shown following the argument of Kauffmann et al. (1993). Let us assume that there is only *one* galaxy per halo, a conservative assumption since we know that the halos of galaxy groups and clusters contain several galaxies. We can then use the Press-Schechter (1974) theory to compute the number density of galaxies as a function of the circular velocity of their surrounding halos, $V_{200}$ (see, for example, eq. 5 of White & Frenk 1991). Assuming that we can infer the rotation velocity of the disk ($V_{opt}$) from $V_{200}$, the Tully-Fisher relation allows us to compute the contribution of galaxies with a given rotation velocity to the mean luminosity density of the universe.

Using the B-band Tully-Fisher relation presented in §5.1, and assuming that the mean luminosity density of the universe is $\mathcal{L}_B = 9.7 \times 10^7 L_\odot$ Mpc$^{-3}$ (Efstathiou et al. 1988b, Loveday et al. 1992), we show in Figure 13 the cumulative fraction of the total luminosity density contributed by galaxies with velocities larger than $V_{opt}$, under the assumption that $V_{200} = V_{opt}$ (solid line). In order to be conservative, we have assumed in this plot that halos with $V_{opt} > 300$ km/s do not contribute to the luminosity density at all, thus excluding groups and clusters of galaxies from the count. As noted by Kauffmann et al. (1993), even with this restriction the cumulative luminosity is too large for $V_{opt} \leq 100$ km/s. This implies that it is impossible to match the Tully-Fisher relation *and* the galaxy luminosity density simultaneously unless our assumption that each halo contains one galaxy is incorrect.

According to the discussion in §5.1, the mean circular velocity of CDM halos, $V_{200}$, should actually be *lower* than $V_{opt}$ in order to account for the shape of the rotation curves of disk galaxies. A simple approximation to the relationship between halo and galaxy circular velocity is

$$V_{200} = F(V_{opt}) = \frac{V_{opt}}{1 + (V_{opt}/300 \, km/s)} \quad (10)$$

(see Figure 11). The result of this identification is shown as a dotted line in Figure 13. Clearly, assigning lower circular velocities to the halos of galaxies of a given luminosity only exacerbates the problem. There is now too much luminosity in the standard CDM cosmogony for $V_{opt} \leq 200$ km/s.

We conclude that this remains a serious problem. Although it can in principle be alleviated by adopting a different cosmogony (for example, an open universe or a universe with cold and hot dark matter components), the most popular forms of these alternative models do not readily overcome the difficulty (Heyl et al. 1995). Another alternative would be to assume that a large number of halos have failed to form galaxies at all, or that the galaxies they host have so far escaped detection. The large numbers of systems being detected in systematic surveys for low surface brightness galaxies suggest that the existence of such a population of galaxies cannot be ruled out (see, *e.g.* Sprayberry et al. 1993, Ferguson & McGaugh 1995).

### 5.5. Core properties of galaxy clusters.

We now examine the consequences of the structure of CDM halos discussed in §4 for X-ray and gravitational lensing observations of galaxy clusters. An ongoing debate concerns the differing estimates of core radius obtained from models of the X-ray emitting gas and from attempts to reproduce the giant arcs observed in many clusters.

The X-ray emitting intracluster medium is often approximated by the hydrostatic, isothermal $\beta$-model (Cavaliere & Fusco-Femiano 1976). The density profile of the X-ray emitting gas is then of the form, $\rho_{ICM} \propto \left(1 + (r/r_{core})^2\right)^{-3\beta/2}$. The core radius, $r_{core}$, is usually a sizeable fraction of the total extent of the emission, and $\beta$ is typically found to be $\sim 0.6$-0.8 (see Jones & Forman 1984). The halo mass profile can then be derived by assuming that the gas is isothermal and in hydrostatic equilibrium; it follows the same law as the gas, with the same core radius and with outer slope parameter $\beta_{DM} = \beta/\beta_T$, where $\beta_T$ measures the ratio of the "temperatures" of the dark matter and of the gas, $\beta_T = \mu m_p \sigma_{DM}^2 / kT_{gas}$ ($\mu m_p$ is the mean molecular weight of the gas and $k$ is Boltzmann's constant). This model is frequently used to interpret X-ray observations and to constrain the dark matter distribution near the center of clusters (Jones & Forman 1984, Edge & Stewart 1991). Cooling flow models with a $\beta$-model mass profile have been successful at explaining a number of observations, including the central X-ray surface brightness "excess" and the drop in central temperature seen in systems with strong cooling flows. The core radius of the dark matter is typically inferred to be in the range $r_c \sim 100\text{-}200$ kpc (Fabian et al. 1991).

On the other hand, such large cores are ruled out by observations of giant arcs, which require core radii of order 20-60 kpc when a $\beta$-model is used to describe the lensing cluster (Grossman & Narayan 1988). This discrepancy is resolved if we assume instead that cluster halos follow the density profile of eq.(3). For the parameters suggested by Figures 7 and 8, halos are sufficiently concentrated to agree with the gravitational lensing constraints. A CDM cluster with mean velocity dispersion $\sim 1000$ km/s placed at $z = 0.3$ can produce giant arcs similar to those seen (Waxman & Miralda-Escudé 1995) yet it can also be consistent with a large core radius in the X-ray emitting gas. Requiring that the ICM be *isothermal* and in hydrostatic equilibrium in the dark matter potential results in a well defined core. This is illustrated in Figure 14, where we plot density profiles for an isothermal gas (dotted line) and for the dark matter (solid line). Radii are given in units of $r_{max}$ and the density units are arbitrary. We assume that the gas and dark matter temperatures are equal, $kT_{gas}/\mu m_p = \sigma_{DM}^2 = GM_{200}/2r_{200}$.

The difference in shape between the gas and the dark matter is due to our assumption that the gas is isothermal. Since the dark matter velocity dispersion drops at radii larger and smaller than $r_{max}$ (see Figures 5 and 6), the gas structure deviates strongly from that of the dark matter at small and large radii. At small radii an isothermal gas develops a well defined core, and at large radii its density drops less rapidly than the dark matter. Such leveling off of the gas profile is not observed in real clusters, indicating that the gas temperature must decrease in the outer regions. Hydrodynamical simulations confirm that this is the case, and indicate that the gas and dark matter distributions in the outer regions are very similar (Navarro, Frenk & White 1995b). The dashed line shows the result of fitting the gas profile with the $\beta$ model mentioned above. In this case, values of $r_{core} = 0.1 \times r_{max}$ and $\beta = 0.7$ give a very good fit to the structure of the gas. For a cluster with a velocity dispersion of 1000 km/s, this implies a gas core radius of about 120 kpc.

Detailed cooling flow models that use the potential corresponding to eq.(3) rather than the $\beta$-model can agree well with observation. A recent analysis by Waxman & Miralda-Escudé (1995) shows that the observational signatures of cooling flows in CDM halos are essentially indistinguishable from those occurring in halos with a true constant density core. We conclude that the apparent discrepancy between X-ray and gravitational lensing estimates of the core radius is a direct result of force-fitting a $\beta$-model to systems whose structure is better described by a profile more similar to that of our CDM halos.

## 6. Conclusions

The main conclusions of this paper can be summarized as follows.

1) The density profiles of CDM halos of all masses can be well fit by an appropriate scaling of a "universal" profile with no free shape parameters. This profile is shallower than isothermal near the center of a halo, and steeper than isothermal in its outer regions.

2) The characteristic overdensities of halos, or equivalently their concentrations, correlate with halo mass in a way which can be interpreted as reflecting the different formation redshifts of halos of differing mass.

3) The observed rotation curves of disk galaxies are compatible with this halo structure provided that the mass-to-light ratio of the disk increases with luminosity. This implies that the halos of bright spirals have masses which correlate only weakly with their luminosity, and may explain why luminosity and dynamics appear uncorrelated in samples of binary galaxies and of satellite/spiral pairs. Disks with rotation velocity in the range 200 to 300 km/s are predicted to have halos with a typical mass of $1.8 \times 10^{12} M_\odot$ within 300 kpc. This agrees well with the masses inferred from the dynamics of observed satellite galaxy samples.

4) CDM halos seem too centrally concentrated to be consistent with observations of the rotation curves

of dwarf irregulars (Moore 1994, Flores & Primack 1994). This may imply that the central regions of dwarf galaxy halos were substantially altered during galaxy formation, for example by sudden baryonic outflows occurring after a burst of star formation (Navarro, Eke & Frenk 1995.)

5) The fact that bright galaxies are surrounded by halos with mean circular velocity *lower* than the observed disk rotation velocity exacerbates the discrepancy between the number of "galaxy" halos predicted in an $\Omega = 1$ universe and the observed number of galaxies.

6) The predicted structure of galaxy clusters is consistent both with X-ray observations of the ICM and with the presence of giant gravitationally lensed arcs. Previous discrepant estimates of the core radius based on these two kinds of data probably result from force-fitting of an inappropriate potential structure. A singular profile such as those favoured by our simulations can be consistent with all current data.

We would like to thank the hospitality of the Institute for Theoretical Physics of the University of California at Santa Barbara, where most of the work presented here was carried out. JFN would also like to thank the hospitality of the Max Planck Institut für Astrophysik in Garching, where this project was started, as well as to acknowledge useful discussions with C. Lacey, N. Katz, and B. Moore. This work has been supported in part by the U.K. PPARC and the National Science Foundation under grant No. PHY94-07194 to the Institute for Theoretical Physics of the University of California at Santa Barbara.

| Label | $L_{box}$ [Mpc] | $1+z_i$ | $h_g$ [kpc] | $M_{200}$ [$10^{12} M_\odot$] | $r_{200}$ [kpc] | $V_{200}$ [km/s] | $N_{200}$ | $r_s/r_{200}$ |
|---|---|---|---|---|---|---|---|---|
| 1 | 3.0 | 44.0 | 1.5 | 0.319 | 177 | 88.0 | 5637 | 0.052 |
| 2 | 3.0 | 44.0 | 1.5 | 0.293 | 172 | 85.6 | 5178 | 0.057 |
| 3 | 3.0 | 44.0 | 1.5 | 0.414 | 193 | 96.1 | 7316 | 0.082 |
| 4 | 3.0 | 44.0 | 1.5 | 0.525 | 209 | 103.9 | 9278 | 0.046 |
| 5 | 6.0 | 22.0 | 3.0 | 2.425 | 348 | 173.1 | 5357 | 0.060 |
| 6 | 6.0 | 22.0 | 3.0 | 2.301 | 342 | 170.1 | 5083 | 0.071 |
| 7 | 6.0 | 22.0 | 3.0 | 3.519 | 394 | 196.0 | 7774 | 0.068 |
| 8 | 6.0 | 22.0 | 3.0 | 2.552 | 354 | 176.1 | 5637 | 0.124 |
| 9 | 12.0 | 11.0 | 6.0 | 28.15 | 788 | 392.0 | 7773 | 0.124 |
| 10 | 12.0 | 11.0 | 6.0 | 20.59 | 710 | 353.2 | 5686 | 0.078 |
| 11 | 12.0 | 11.0 | 6.0 | 29.67 | 802 | 398.9 | 8193 | 0.131 |
| 12 | 12.0 | 11.0 | 6.0 | 22.01 | 726 | 361.1 | 6078 | 0.088 |
| 13 | 12.0 | 11.0 | 6.0 | 26.16 | 769 | 382.5 | 7224 | 0.077 |
| 14 | 12.0 | 11.0 | 6.0 | 22.65 | 733 | 364.6 | 6254 | 0.065 |
| 15 | 18.0 | 5.5 | 9.0 | 102.68 | 1213 | 603.4 | 8401 | 0.110 |
| 16 | 24.0 | 5.5 | 12.0 | 224.35 | 1574 | 783.0 | 7744 | 0.121 |
| 17 | 40.0 | 5.5 | 20.0 | 1109.9 | 2682 | 1334 | 8274 | 0.151 |
| 18 | 48.0 | 5.5 | 24.0 | 1931.5 | 3226 | 1605 | 8334 | 0.188 |
| 19 | 58.0 | 5.5 | 30.0 | 3009.7 | 3740 | 1861 | 7360 | 0.143 |

Table 1: Parameters of the numerical experiments.

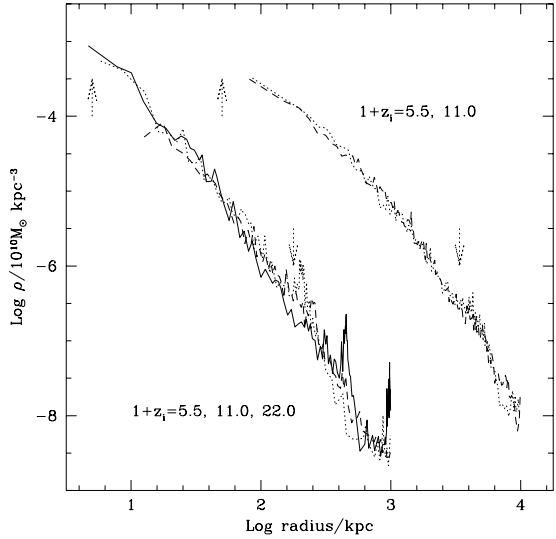 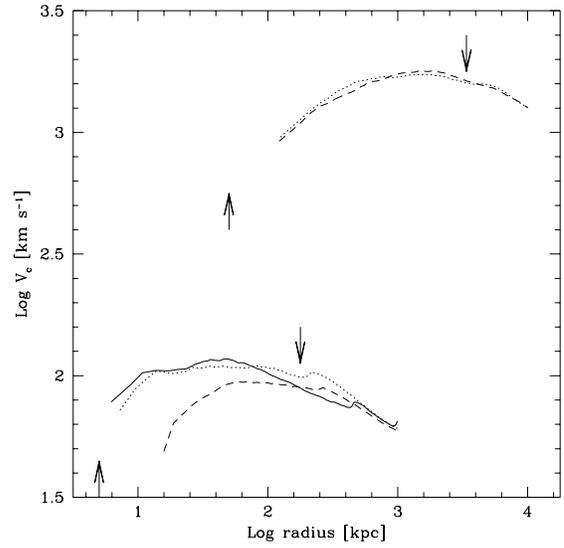

Fig. 1a.— The effects of varying the initial redshift on the density profile of simulated halos. The masses of the two halos shown are $\sim 10^{11}$ and $\sim 10^{15} M_\odot$. The gravitational softening and the virial radius are indicated with upward and downward pointing arrows, respectively. Solid, dotted, and dashed lines correspond to $1 + z_i = 22.0$, $11.0$, and $5.5$, respectively.

Fig. 1b.— Circular velocity profiles for the halos shown in Fig. 1a. Arrows are also as in Fig. 1a. For $1 + z_i = 5.5$, the circular velocity near the center of the small halo is substantially underestimated. For larger $z_i$, the models converge to a unique profile.

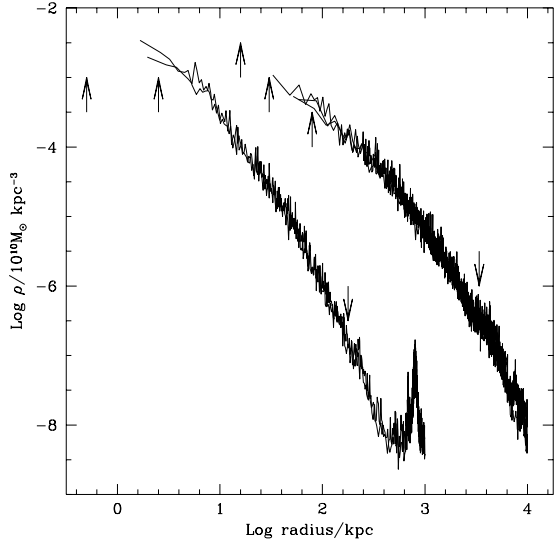

Fig. 2a.— Density profiles of the same halos shown in Figure 1 for different choices of the gravitational softening, $h_g$. Upward-pointing arrows indicate the softening values. Downward pointing arrows indicate the virial radius. Varying the softening by a factor of five has no significant effect on the structure of the halo beyond one gravitational softening length.

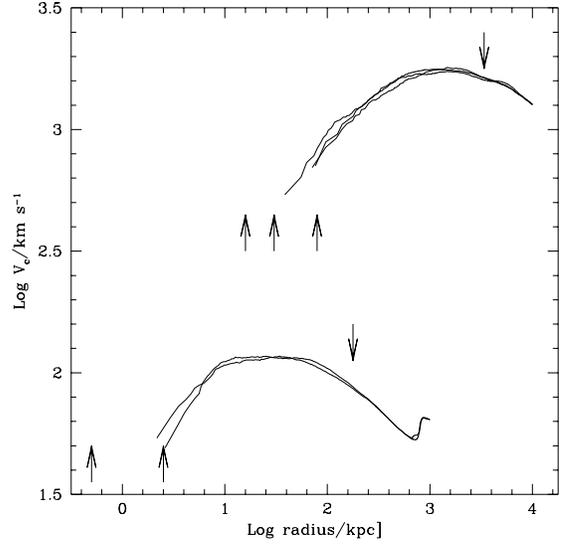

Fig. 2b.— Circular velocity profiles for the halos shown in Figure 2a. Arrows are also as in Figure 2a. Differing gravitational softening lengths have no significant effect on the structure of the halo on regions larger than $h_g$.

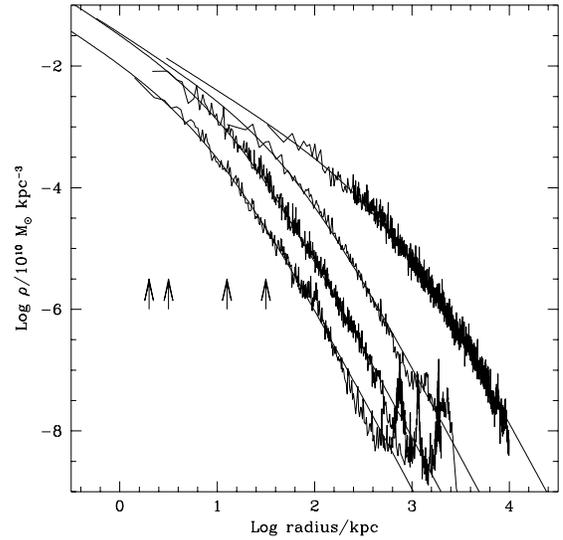

Fig. 3.— Density profiles of four halos spanning four orders of magnitude in mass. The arrows indicate the gravitational softening, $h_g$, of each simulation. Also shown are fits from eq.3. The fits are good over two decades in radius, approximately from $h_g$ out to the virial radius of each system.

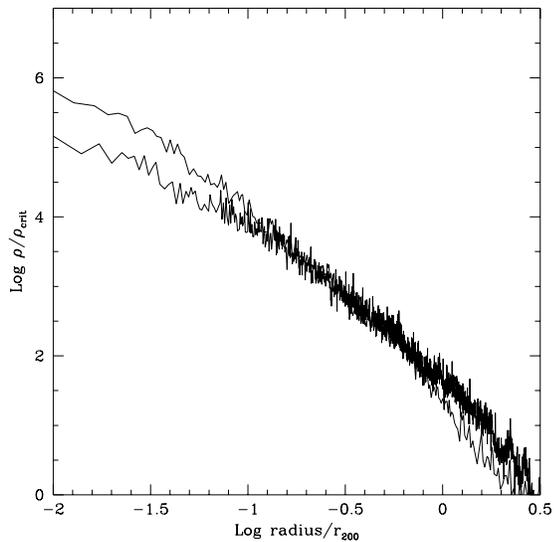

Fig. 4.— Scaled density profiles of the most and least massive halos shown in Figure 3. The large halo is less centrally concentrated than the less massive system.

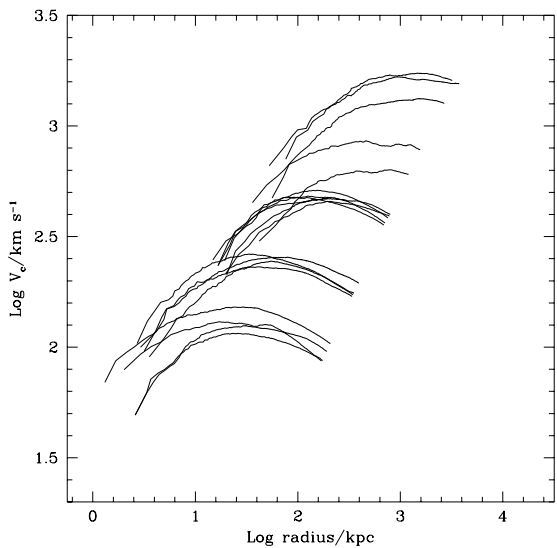

Fig. 5.— Circular velocity profiles of all 19 halos. The profiles are truncated at the virial radius, $r_{200}$. The gravitational softening is about $10^{-2} \times r_{200}$. Note that all profiles have the same shape.

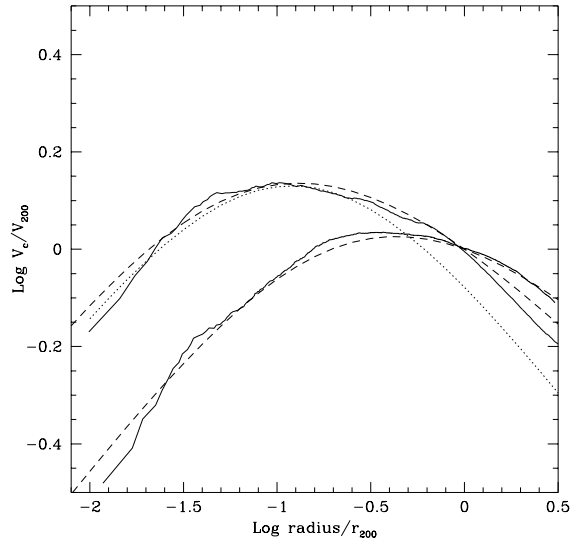

Fig. 6.— Scaled circular velocity profiles of two halos, one of the largest and one of the smallest in our sample (solid curves). The dashed lines are fits with eq.(5). The dotted line is a fit to the low mass system using a Hernquist model (see eq.6). Note that the Hernquist model underestimates the halo circular velocity at $r_{200}$.

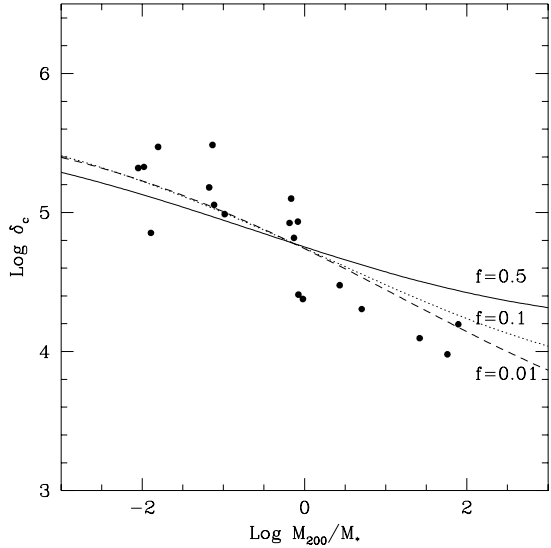

Fig. 7.— The characteristic overdensity $\delta_c$ as a function of the mass of the halo. The curves show the mass-overdensity relation predicted from the formation times of halos (see text). All curves are normalized so that they cross at $M_{200} = M_\star$.

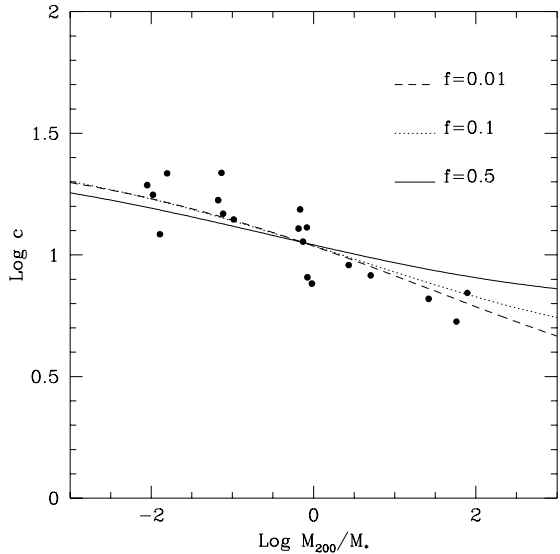

Fig. 8.— The concentration $c$ as a function of the mass of the halo. The curves show the mass-concentration relation predicted from the formation times of halos. All curves are as in Figure 7, and have been normalized so that they cross at $M_{200} = M_\star$.

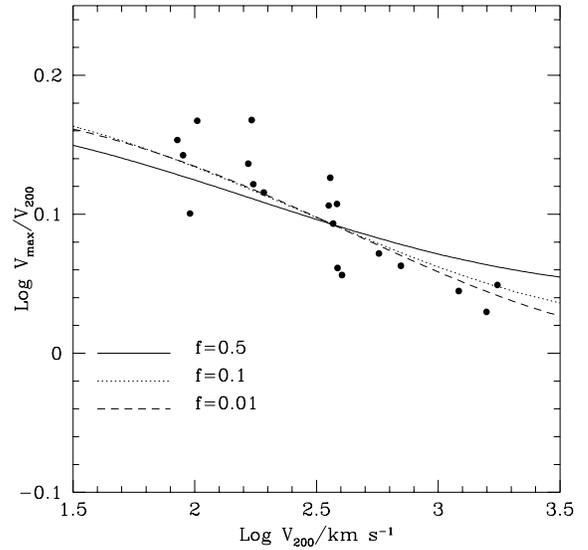

Fig. 9.— The ratio between the maximum circular velocity of a halo and the value at $r_{200}$ as a function of the circular velocity at the virial radius. Low mass systems are more concentrated than more massive ones. The maximum velocity can be up to twice the value at $r_{200}$ for a small galaxy halo.

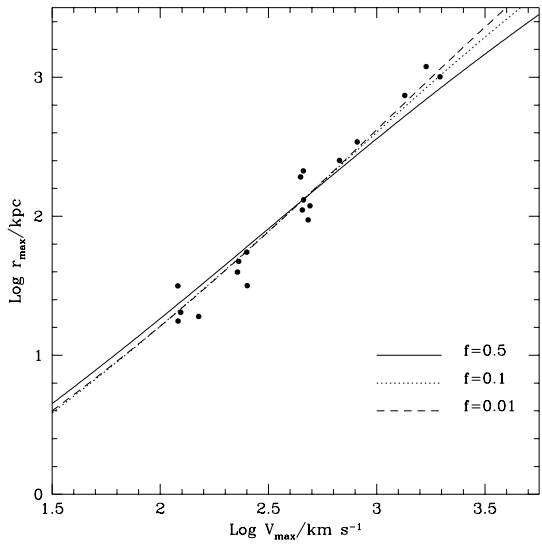
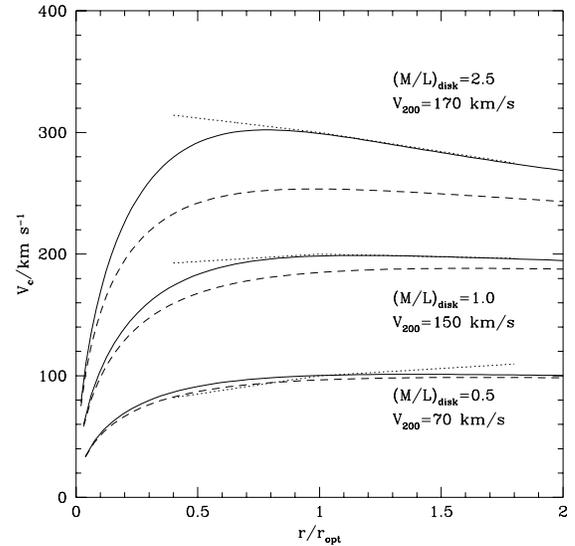

Fig. 10.— The maximum circular velocity as a function of the radius at which it is attained in halos of different mass. Note that a halo with $V_{max} = 220$ km/s has a rising rotation curve that extends out to about 50 kpc, well beyond the luminous radius of a galaxy like the Milky Way.

Fig. 11.— Rotation curves of disk+halo systems (solid lines) with parameters chosen to match the observational data of Persic & Salucci (1995) (dotted lines). The dashed lines indicate the contribution of the dark halo. Note that the disk mass-to-light ratio increases as a function of mass, and that the halo contribution is less important in bright galaxies.

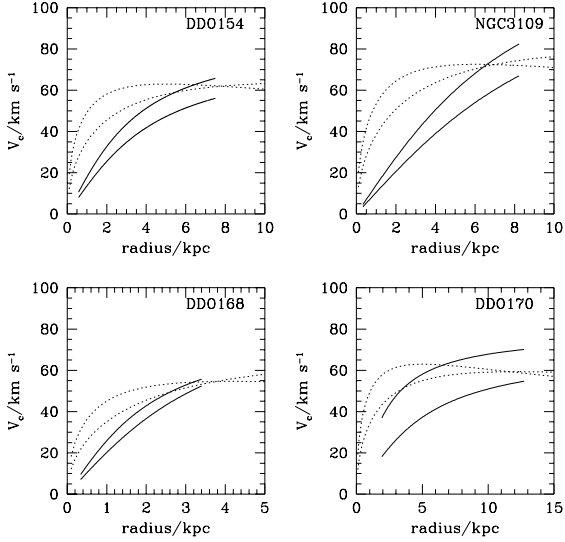
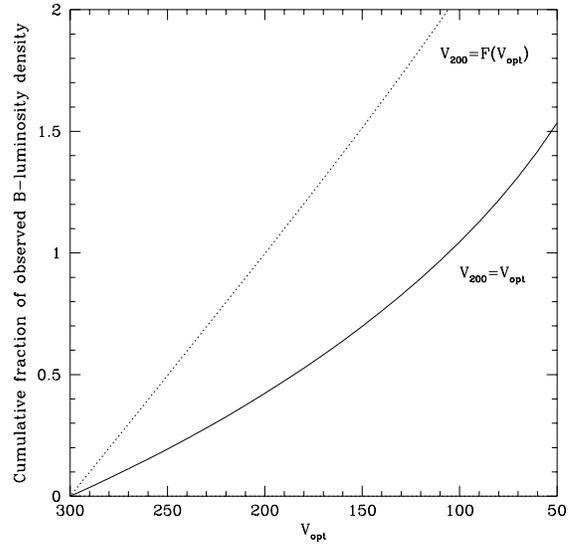

Fig. 12.— The circular velocities of CDM halos (dotted lines) compared with the halo contribution to the rotation curve of four dwarf galaxies (solid lines). The solid lines encompass the likely contribution of the halo, and correspond to the "maximal" and "minimal" disk hypotheses. CDM halos seem to be significantly more concentrated than allowed by observations.

Fig. 13.— The cumulative B-band luminosity density in galaxies with optical rotation velocities larger than $V_{opt}$ as predicted by our standard CDM model. We use the Tully-Fisher relation, and assume that the abundance of galaxies with rotation velocity $V_{opt}$ matches that of halos with $V_{200} = V_{opt}$ (solid line), or $V_{200} = F(V_{opt})$ as given in the text (dotted line).

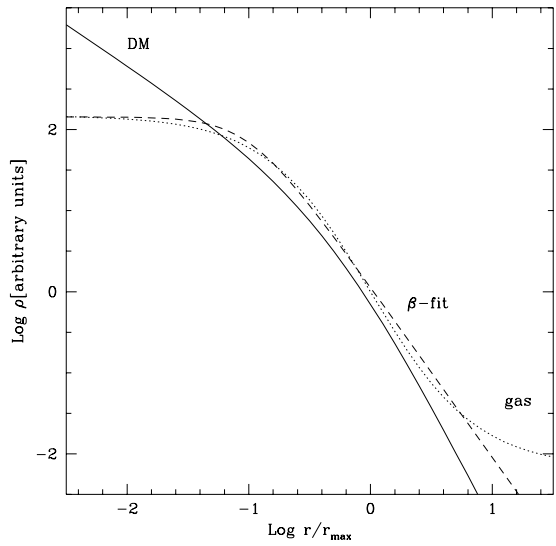

Fig. 14.— The density profile of an isothermal gas (dotted line) in hydrostatic equilibrium within a CDM halo whose structure is given by eq.(3) (solid line). The dashed line indicates a fit using the $\beta$ model. The parameters $r_c = 0.1\,r_{max}$ and $\beta = 0.7$ give an excellent $\beta$-model fit to the gas profile.